# Magnetism at iridate/manganite interface: influence of strong spin-orbit interaction.

G.A. Ovsyannikov[1], T.A. Shaikhulov[1], V.V. Demidov[1], K.L.Stankevich[1], Yu. Khaydukov[2,3], N.V. Andreev[4]

[1]Kotel'nikov IRE RAS, Mokhovaya str., Moscow, Russia
[2]Max Planck Institute for Solid State Research, Stuttgart, Germany
[3]Max Planck Society Outstation at the MLZ, Garching, Germany
[4]National University of Science & Technology (MISIS), Lenin av., Moscow, Russia

Abstract

The complex investigation of dc transport and magnetic properties of the epitaxial manganite/iridate heterostructure was carried out by mean of X-ray (XRD), dc resistance measurements, ferromagnetic resonance (FMR) and polarized neutron reflectivity (PNR). Epitaxial growth of the heterostructure proceeded according to the "cube-to-cube" mechanism with the small lattice turn. The dc measurement indicates the presence of a conduction channel at the iridate/manganite interface due to the charge leakage from iridate that makes it hole doped, while the manganite side becomes electron doped. This is confirmed by the first principles calculations based on density functional theory [Sayantika Bhowal, and Sashi Satpathy AIP Conference Proceedings 2005, 020007 (2018)] that show the charge transfer at the interface from the half-filled spin-orbit entangled $J_{eff} = 1/2$ state of the iridate to the empty $e^{\uparrow g}$ states of manganite. The neutron scattering data show the turn of magnetization vector of the heterostructure (mainly manganite) on 26 degree closer to the external field with reducing temperature down to 10K. Additional ferromagnetic state appearing at T<100K indicate on emergence of ferromagnetism in the thin (10 nm) paramagnetic SIO film close to the interface. We have measured the dc voltage aroused on the SIO film caused by spin pumping and the anisotropic magnetoresistance in the heterostructure.

1. Introduction

Transition metal oxides (TMOs) which differ from binary oxides like $SiO_2$ to more complex compounds contents are nowadays a subject of intense activities in condensed matter physics. 3d TMO have various functionalities, including ferromagnetism caused by the presence of strong electron-electron correlation (energy U) [1, 2]. However, the spin-orbit interaction (with energy $E_{SO}$ ) is usually weak or insignificant in 3d-TMOs. On the other side 5d -TMOs induce considerable interest due to the occurrence of a strong spin-orbit interaction, which coexists along with the electron-electron interaction. The combination of spin-orbit interaction and electron-electron interaction as predicted theoretically [3, 4.], makes it possible to realize several new quantum states of matter, such as the topological Mott insulator [5, 6], the quantum spin Hall effect, the quantum anomalous Hall effect [7, 8, 9], the Weyl semimetal [10] and even a high-temperature superconductor [11, 12]. The contact between 3d and 5d TMOs provides a unique interface, in which the existence and interaction of electron-electron and spin-orbit interactions is possible, unlike the well-studied 3d/3d TMO-interfaces [13-17]. Reconstructed magnetic anisotropy and strong spin orbital interaction indicate that the 3d/5d interfaces are objects for observing magnetic texture and topological phenomena [18]. At the interface of such a material with a ferromagnetic, a violation of topological symmetry in the region of the interface and the occurrence of a gap in the excitation spectrum can occur, which in turn can lead to rather strong magnetoelectric effects. These interfaces provide the ideal candidates to search for novel magnetic textures and topological phenomena. Moreover, due to the inherent mixture of the spin and orbital degrees of freedom in the 5$d$ TMOs, these heterostructures also provide the potential pathways to achieve the electric field control of magnetism through the mechanisms that have not been demonstrated in 3$d$/3$d$ heterostructures. Current research is still in the early stage and limited to a few systems, and more systematic investigations are highly desired to fully unravel the unique role of 5$d$ TMOs.

Iridate $SrIrO_3$ single crystals have a slightly distorted $SrRuO_3$-type orthorhombic structure (a = 0.560 nm, b = 0.558 nm, c =0. 789nm) of the Pbnm space group [19]. Thin $SrIrO_3$ epitaxial films form a perovskite structure due to the intensity the interaction with the substrate during the film growth. Such films can be described as a distorted pseudo-cubic with a constant 0. 396nm [20-28]. Due to the crystal structure similar to manganites, the epitaxial films of the $SrIrO_3$ iridate can be an excellent component for the growth of heterostructures with manganites. The strong interaction of spin and orbital degrees of freedom leads to the fact that the low-energy state of 5d electrons of the $Ir4^+$ state is half full ($J_{eff}$ = 1/2 state) and the energy spectrum differs significantly from 3d manganites [29]. In 5d transition metals, $E_{SO}$≈0.4 V is several times higher than $E_{SO}$ 3d transition metals and is comparable with the energy of electron correlations U ~ 0.5 eV. Experimentally, $SrIrO_3$ is a paramagnetic metal which turns into a paramagnetic insulator below a transition temperature $T_{MI}$ = 44 K [30].

Good crystalline correspondence between $SrIrO_3$ epitaxial films and other perovskites allows the creation of $SrIrO_3/La_{1-x}Sr_xMnO_3$ superlattices with different x values [15, 31.] and $SrIrO_3/SrTiO_3$ [14]. In the $SrMnO_3/SrIrO_3$ superlattice [17] interface forms a non-polar boundary, the occurrence of magnetism (the emergence of the ferromagnetism) in antiferromagitic $SrMnO_3$, caused by hybridization of Mn and Ir orbitals was observed. As the thickness of the SrIrO$_3$ layer in the superlattice changes, the axis of easy magnetization of the manganite layer is rotated between the crystallographic directions: (110) La$_{0.7}$Sr$_{0.3}$MnO$_3$ and (001) La$_{0.7}$Sr$_{0.3}$MnO$_3$ [15, 31]. Recently the transport properties and ferromagnetic resonance spectra of heterosructure La0.7Sr0.3MnO3 /SrIrO3 was investigated were investigated [32-34]. The parameters of heterostructure are compared with the properties of individual iridate and manganite films. As temperature decreases, the ferromagnetic resonance line width increases and the resonance field decreases[32,33]. Influence of spin –pumping on magnetic damping was observed [34].

In this paper, we present the results of the growth of an epitaxial heterostructure of a iridate with strong spin-orbit interaction (SrIrO$_3$) and ferromagnetic manganite (La$_{0.7}$Sr$_{0.3}$MnO$_3$), which has a Curie temperature above room temperature and we give the results of electrical, magnetic and neutron measurements of the heterostructure. The remaining parts of the paper are organized as follows. The heterostructure fabrication and X-ray data are presented in Sec. 2. Its indicated on the growth of the heterostructure proceeded according to the "cube-to-cube" mechanism with the small lattice turn. This is follows by dc measurements of resistance for simple films deposited on the substrate and heterostructures (Sec 3). The charge transport at the interface in the heterostructure differs significantly from both transport in individual films and simple metals. In Sec. 4 the data measured by SQUID and neutron polarized neutron reflectivity are presented. Temperature dependence of the saturation magnetization of the heterostructure and neutron data well correspond to the mean-field approximation. Neutron experiment show the turn of magnetic vector of the heterostructure on 26 degrees. Ferromagnetic resonance was measured for the heterostructure with significantly low thickness of the films in the heterostructure. Additional ferromagnetic ordering was observed in the heterostructure that have different magnetic anisotropy then the one for the base La$_{0.7}$Sr$_{0.3}$MnO$_3$ film (Sec. 5). Finally in Sec. 6 spin current aroused for ferromagnetic resonance across the interface in heterostructure was compare with voltage induced by anisotropic magnetoresistance. A summary of the paper in presented at Sec.7. The method for determination of magnetic parameters of the heterostructure using angular dependence of FMR spectrum is given in Appendix.

2. Heterostructure fabrication and X-ray data

Heterostructures were obtained by magnetron sputtering on a neodymium gallium monocrystal substrate with orientation (110) NdGaO$_3$ (NGO) at a temperature T = 820C and an oxygen pressure of 0.7 mbar for lanthanum stronsium manganite La$_{0.7}$Sr$_{0.3}$MnO$_3$ (LSMO) film and T = 770C and a pressure of 0.3 mbar for iridate SrIrO$_3$ (SIO). After deposition, the films were cooled in situ to 600C in 1 atm oxygen for 10 min and then were cooled down to room temperature in 20 min. Film thicknesses varied from 5 to 50 nm. We used the deposition time to control the SIO thickness after measuring thick of the film using alfa-Step technique [32].

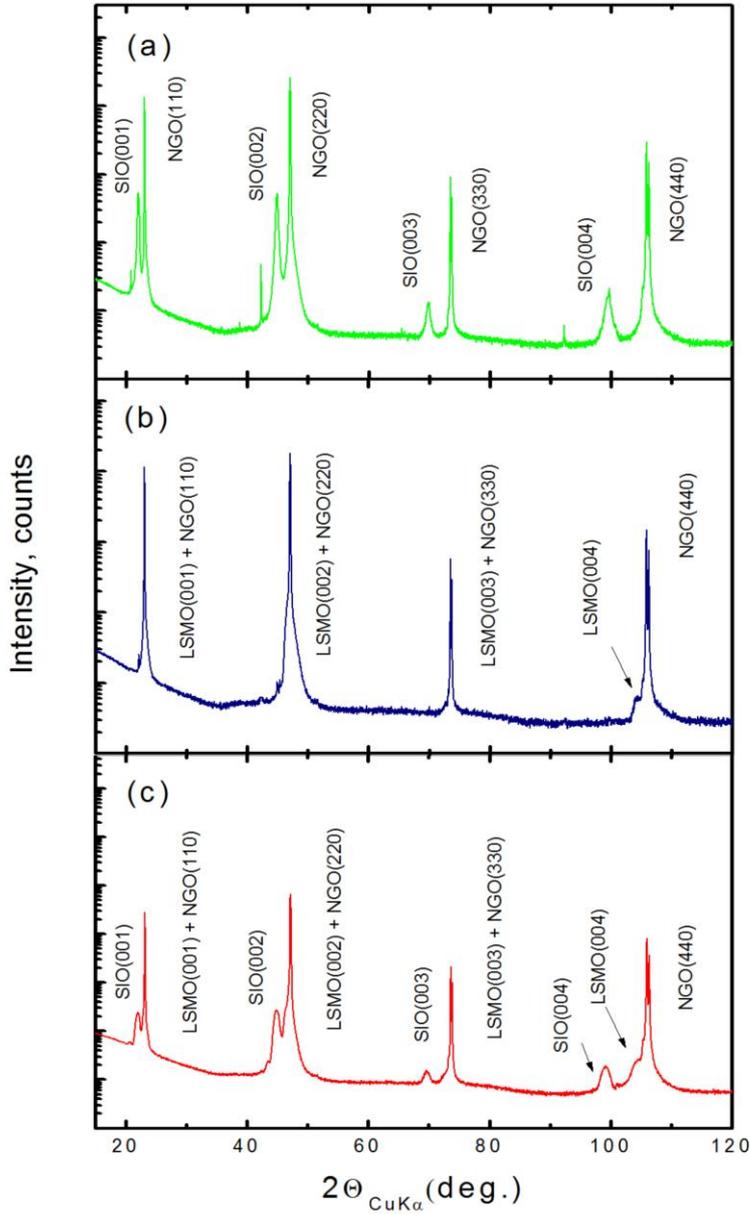

Fig.1. X-ray 2θ/ω scan for autonomous (single layer) a)-SIO film alone and b)- LSMO film, as well as c)- SIO/LSMO heterostructure, all deposited on a (110) NGO substrate

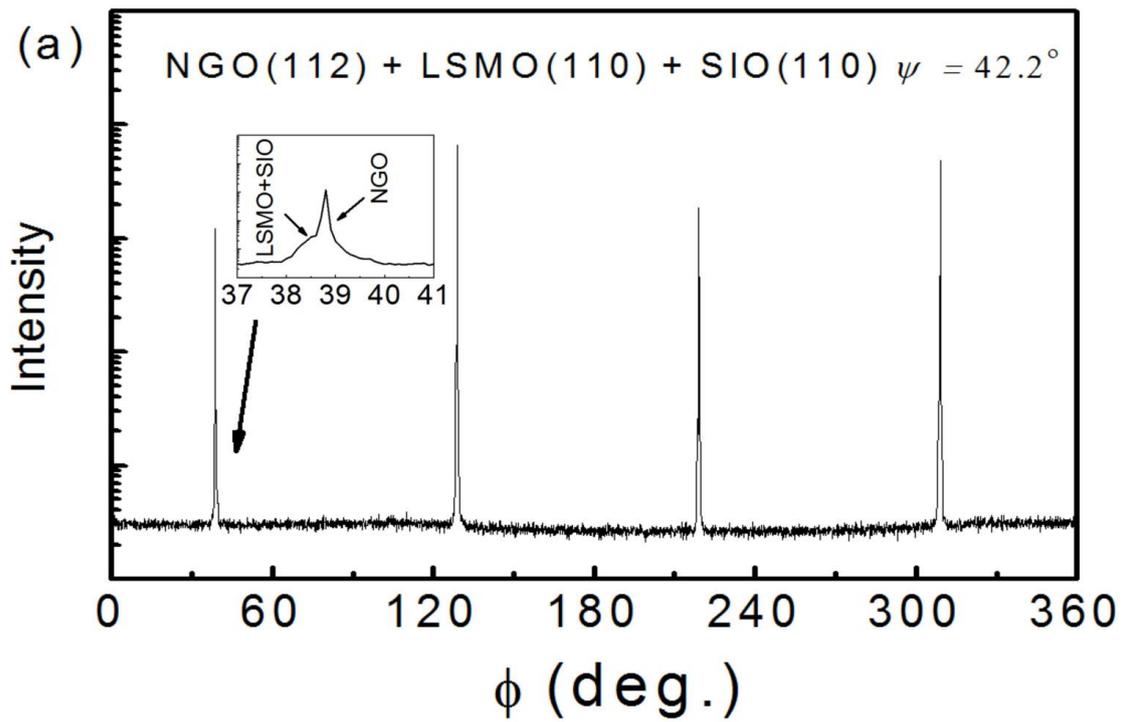

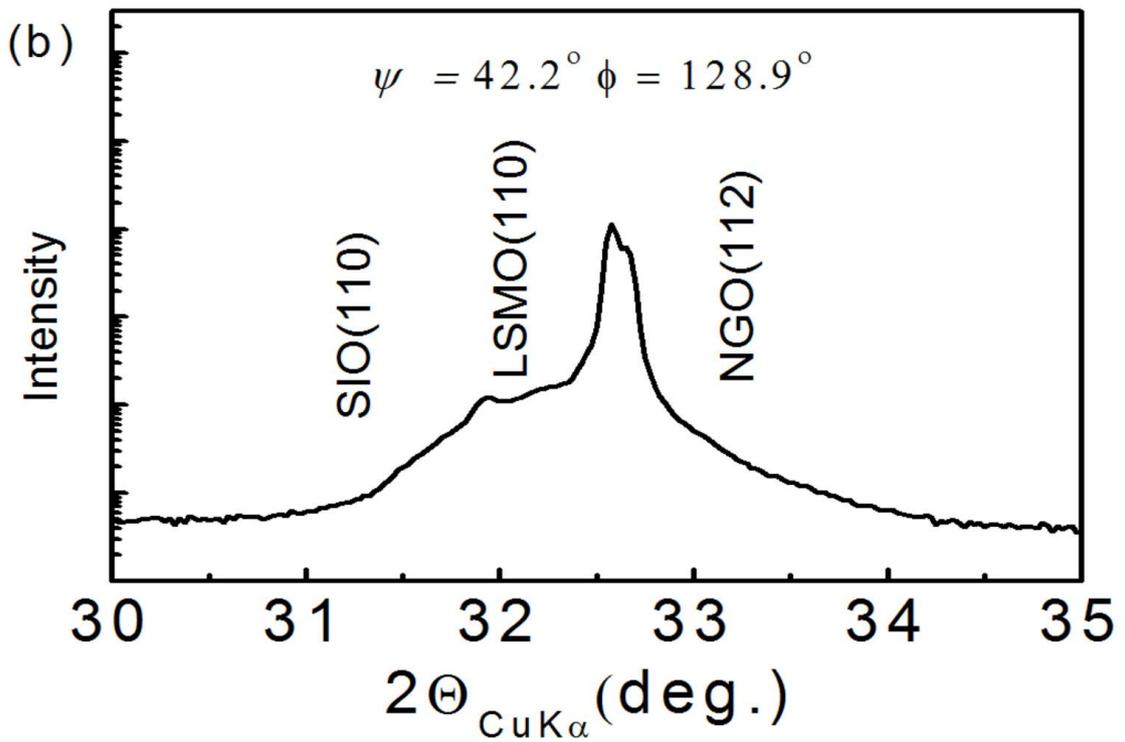

Fig.2a-XRD φ-scan at ψ= 42.2 for the SIO/LSMO heterostructure tilted to (112) NGO, inset shows enchantment of single peak;b- 2θ-ω scan of the heterostructure for tilted configuration ω = 42.2° and φ=128.9°

The crystal structure and microstructure of SIO thin films and heterostructures were characterized by x-ray diffraction (Bruker Discover VIII using CuKa radiation). Figure 1(a) shows the $2\theta$-$\omega$ scan of the autonomous thin film SIO. The observed peaks correspond to multiple reflections from the plane (110) NGO substrate and from the (001)SIO film (the pseudo-cubic notation). This suggests that the film is out-of-plane orientated (001)SIO||(110)NGO. The similar pattern can be seen for the autonomous thin film LSMO (Figure 1(b)). If we describe the LSMO lattice as a pseudo-cubic with the parameter a = 0.389 nm [35] then visible peaks reveal that the film also is out-of-plane-oriented (001)LSMO||(110)NGO. The $2\theta$-$\omega$ scan of heterostructure SIO/LSMO deposited on NGO substrate is the superposition of autonomous single-layer films scans (Fig. 1(c)). The out-of-plane lattice parameter in the LSMO film does not change significantly with growth in the heterostructure remaining the same as for an autonomous film $d_{LSMO}$=0.388 nm. A slight change is observed in the out-of-plane lattice parameter for the autonomous SIO film from $d_{SIO}^A$ = 0.403 nm in the autonomous case to $d_{SIO}$ = 0.404 nm for the heterostructure.

Figure 2a shows XRD $\phi$-scan at tilt angle $\psi$ = 42.2° and $2\theta$ =38.5° angle for the (112) NGO plane for the heterostructure SIO/LSMO. In addition to the four strong peaks from the substrate spaced at almost 90 degrees (weak orthorhombic of the substrate NGO) one can observe reflections from the (110)SIO and (110)LSMO planes, peaks coincide and are displaced from the substrate peaks by approximately 0.3° (see insert Fig.2b). The corresponding $2\theta$-$\omega$ scan with $\psi$ = 42.2° and $\phi$ = 128.9° is presented in Fig. 2(b). Thus we can conclude that the growth of the heterostructure proceeded according to the "cube-to-cube" mechanism with the small lattice turn. The epitaxial relationships are as folllowing: (001)SIO||(001)LSMO||(110)NGO and [100]SIO||[100]LSMO||[001]NGO. Narrow rocking curve (FHMW = 0.1–0.12°) indicate the high quality of the films.

3. DC transport of the heterostructure

The electrical properties of the films and heterostructure were measured by the four-probe method current-in-plane sheet resistance using Mangomery technique [36]. The sheet resistance of the film or heterostructure is measured in this case. To compare the transport between the autonomous (single layer) films and the heterostructure we plotted the temperature dependent resistance curves (Fig. 3).

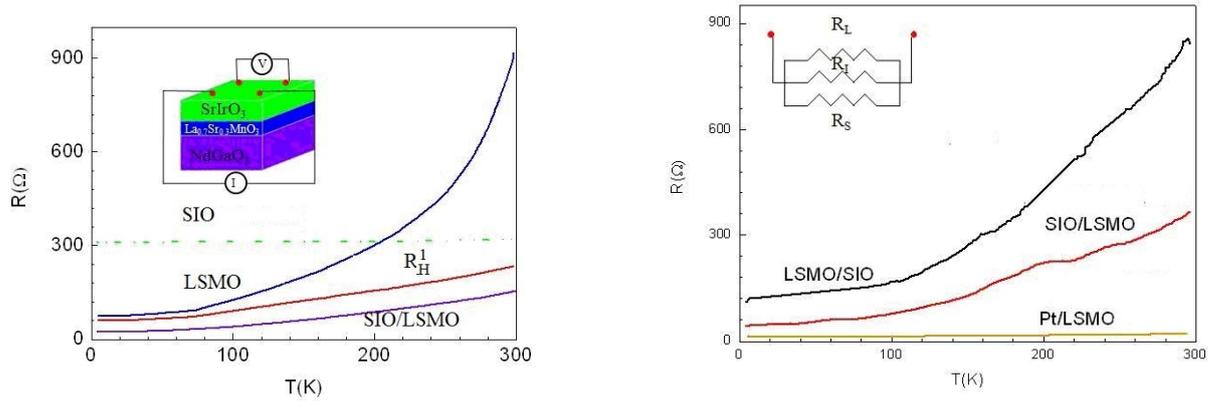

Fig.3.a-Temperature dependence of the resistance of the SIO and LSMO films with thickness 10 nm and 12 nm correspondingly, as well as the heterostructures SIO/LSMO with thicknesses 12 nm and 10 nm for SIO and LSMO film respectively. $R_H^1$ is calculated temperature dependence of the SIO/LSMO heterostructure resistance assuming parallel connection of the film LSMO and SIO resistances. The circuit for resistance measurements is shown on inset. b - Temperature dependence of interface resistance $R_I$ for following heterostructures: SIO(10nm)/LSMO(12nm), LSMO(15nm)/SIO(10nm) and Pt(10nm)/LSMO(20nm). The film thicknesses are indicated in brakes. Equivalent circuit is shown on inset. $R_S$ and $R_L$ are resistances of SIO and LSMO films correspondingly

Fig 3a shows the temperature dependence of the resistance of the epitaxial LSMO and SIO films grown on insulating NGO substrate. LSMO films have metallic behavior at temperatures below temperature of metal-insulator transition ($T_M$) consistent with previous reported behavior of epitaxial LSMO film on these substrates [34] In manganites, a resistivity maximum at $T=T_M$ commonly is observed at near the Curie temperature [34, 37]. The temperature-dependent resistance of the SIO film is also reduced with lowing temperature but no so significant compare with LSMO films (Fig.3a) [21, 28, 38].

To assist in the interpretation of the resistance data, we modeled the resistance ($R_H^1$) of the SIO /LSMO heterostracture as a parallel connection of the resistance of the upper layer of the SIO film $R_S$ and the parallel-connected resistances of the bottom LSMO layer $R_L$. $R_H^1 = R_S R_L /(R_S + R_L)$. It is larger then measured resistance of heterostructure ($R_H$). The presence of the interface resistance connected in series with $R_L$ increased the difference between $R_H^1$ and $R_H$ [39]. Possible solution is taking into account the parallel connected of interface resistance $R_I$ as shown on inset to Fig.3b. Using obtained sheet resistance of interface SIO/LSMO equal to $R_I = \rho_I/d_S$ we get $\rho_I = 8 \cdot 10^{-6}$ $\Omega \cdot cm$ for low temperature suppose the thickness of interface is equal to 1 nm. So small resistivity of the interface indicate on the possibility to exist the layer of electronic gas with high mobility [1, 40].

Oxides of transition metals due to the presence of strong electron correlations are significantly different from simple metals. The presence of a large number of degrees of freedom — spin, charge, lattice, and orbital — leads to the complexity of the behavior of these materials, especially in the interface region. The charge transport at the interface in the heterostructure differs significantly from both transport in individual films and simple metals.

While the electronic properties of the 3$d$ TMO is governed by the strong Coulomb interaction, in the 5$d$TMO, the spin-orbit interaction is a dominating. It was shown theoretically that in the ballistic regime in a two-layer system consisting of a magnetic insulator and an adjacent non-magnetic metal the interfacial current appear due to spin-orbit interaction in the metal layer near the interface[41]. The induced current could strong influence on the interface resistance that observed in our case (Fig. 3b.)

The first principles calculations based on density functional theory have been performed in [42]. It was shown that the charge transfer at the interface from the half-filled spin-orbit entangled $J_{eff}$ = 1/2 state of the iridate to the empty $e^{\uparrow g}$ states of iridate. The charge leakage from iridate makes it hole doped, while the manganite side becomes electron doped. The doped carriers make both sides metallic. Approximately the same charge transfer is obtained if one integrates the partial density of states for the Ir and Mn atoms. According to the calculation [40] the charge transfer there is a transfer of 0.06 electrons per interface Ir atom from the iridate to the manganite side. This leads to the charge transfer across the interface, which is comparable to the density of the 2DEG in the well-studied polar interface of LaAlO3 and SrTiO3 (see for example [43]).

4. Neutron scattering

For the neutron experiment we prepared a heterostructure having properties of neutron magnetic waveguide [44]. This design allows us to get additional sensitivity to the appearance of small magnetic moment in the SIO layer. To make this design we have covered LSMO/SIO structure by a gold layer (see Fig. 4a) and increase the thickness of the films in heterostructure. Fig. 4b show the hysteresis loops of resulted Au(70nm)/LSMO(50nm)/SIO(44nm) structure measured by SQUID at T=300K and T=100K. The loops were corrected on paramagnetic background of the NGO substrate. The paramagnetic signal is increasing rapidly with cooling and leads to the elevated error bar for T<100K. The temperature dependence of saturation magnetization compared with neutron data is presented below (Fig.7b).

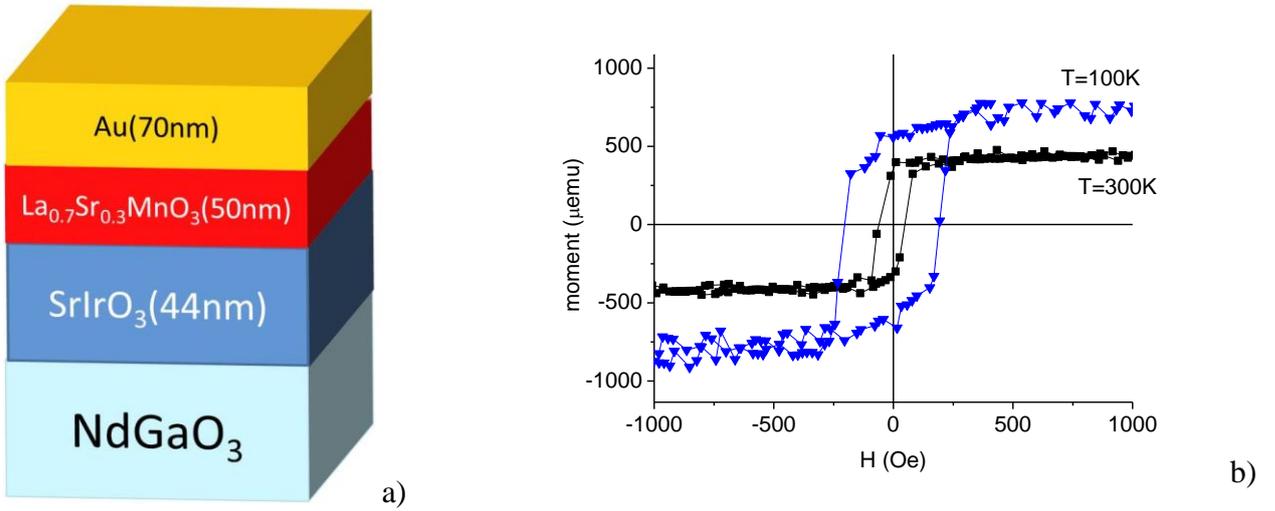

Fig. 4. a- The sketch of the structure measured by SQUID and neutron polarized neutron reflectivity. b-The hysteresis loops of the heterostructure measured at T=300K and T=100K background due paramagnetic of the substrate NdGaO3 was corrected.

The polarized neutron reflectivity (PNR) experiments were conducted on the angle-dispersive reflectometer NREX ($\lambda$ = 0.428 nm) at the research reactor FRM-II (Garching, Germany). The polarized neutron beam (with polarization 99.99%) was incident on a sample at grazing angles $\theta$ = (0.15°–1°). The polarization of the reflected beam was analyzed by an analyzer with efficiency 99.2%. In the experiment we applied small external magnetic field H = 5Oe in the plane of the sample and with 5 degree accuracy along one of the edges of the substrate (Fig.5a). At a fixed temperature, we measured four reflectivity curves for fixed positions of polarizer and analezer $R^{++}$, $R^{--}$, $R^{+-}$, and $R^{-+}$ as a function of momentum transfer Q=4$\pi$ sin$\theta$/$\lambda$ (Fig.5 b,c). The non-spin-flip (NSF) reflectivities $R^{++}$, $R^{-}$ are sensitive to the sum and the difference of the nuclear scattering length density (SLD) profile and the in-plane magnetization component $M_{\parallel}$ collinear with the external magnetic filed H. In order to separate magnetic signal from nuclear it is convenient to analyze spin asymmetry S$\equiv$($R^{++}$- $R^{--}$)/($R^{++}$+ $R^{--}$) which is proportional to $M_{\parallel}$. The spin-flip (SF) reflectivities $R^{+-}$, $R^{-+}$ in turn are sensitive to square of non-collinear to H in-plane component of magnetization $M_{\perp}$. In majority of PNR experiments, including ours, $R^{+-}$(Q)=$R^{-+}$(Q). In this regard, for analysis we averaged spin-flip reflectivities $R^{SF}\equiv[R^{+-}+ R^{-+}]/2$ to improve statistical accuracy [45-50].

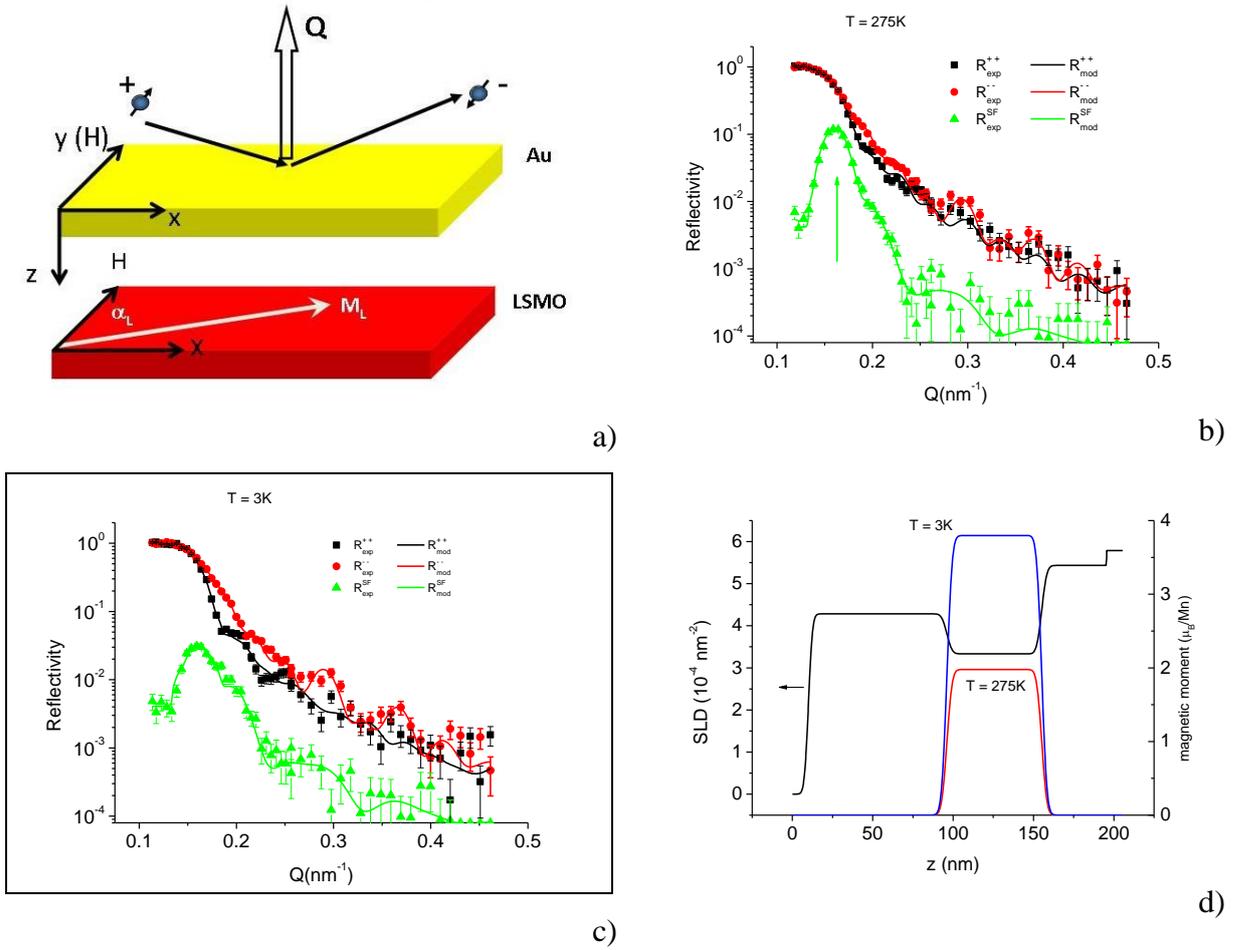

Fig. 5a-The topology of neutron experiment. α is the angle between the magnetization direction of LSMO film and external magnetic field. Experimental (dots) and model (lines) polarized neutron reflectivity measured at b-T=275K, c- T=3K. Arrow indicates the critical angle. d- the depth profiles of nuclear SLD at T=275K and magnetics moment both at T=275K and T=3K .

Fig.5b and Fig.5c shows the PNR data measured at T=275K and T=3K correspondingly. We observed strong SF scattering with the resonance peak in the vicinity of critical momentum transfer $Q_{cr}$= 0.16 $nm^{-1}$ having intensity of $R^{SF}(Q_{WG})$ =12% (arrow in Fig.5b). At the same time we observed non-zero spin asymmetry with maximum of S = -20% slightly above $Q_{cr}$. These curves can be fitted with the nuclear SLD profile depicted in the inset to Fig. 6c and LSMO magnetic moment that changed with reduce temperature. At T=300K it equal to 2 $\mu_B$/Mn tilted on the angle on angle α=38° to the direction of external field [453, 47-49].

The Fig.6a shows that suppression of integrated spin-flip scattering (SF) and increase of averaged spin-asymmetry (SA) takes place systematically below 150K. After cooling of the sample down to 3K we observed 3 times decreased intensity of SF scattering accompanied by a 2 times increased SA (Fig. 6a). This behavior can naturally be explained by the turn of magnetization vector of LSMO closer to the external field. The temperature dependence of the saturation moment of the saturation moment well fitted within the mean field theory with bulk Curie temperature $T_m=340K$ (Fig.6b) [44]. Quantitatively we can describe the data at 3K by LSMO magnetic moment $3.7\mu_B$/Mn turned at $\alpha = 26°$ (Fig. 7b).

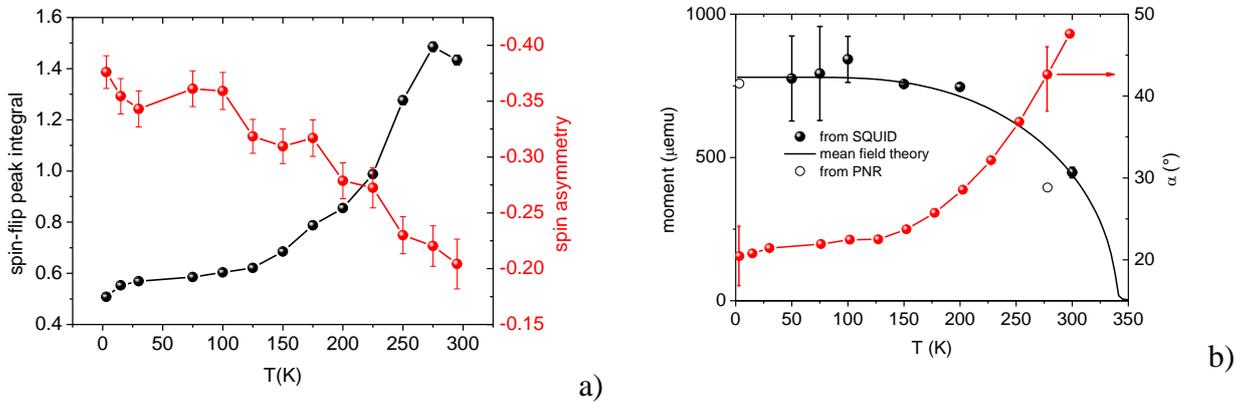

Fig.6a- The temperature dependence of integrated spin-flip scattering (black circles) and averaged spin-asymmetry (red circles).b- The saturation magnetization versus temperature (black circles) and mean-field approximation with Curie temperature $T_{CU}=340K$ (solid line).The neutron data are also shown on the curve(open circles).

We also tried another model at which magnetization of LSMO was fixed to 3.7 $\mu_B$/Mn and magnetization of SIO varied. Since the whole structure was designed as a magnetic waveguide sensitive to the appearance of magnetic moment in SIO we indeed can describe suppression of the SF peak by small positive magnetization (10% of LSMO moment) in whole SIO layer. However such small distortion of magnetic contrast can not describe increase of spin asymmetry. On the other hand presence of magnetic moment at the interface SIO/LSMO layer (~1nm thickness) is beyond sensitivity of PNR.

The density-functional results [40] show that a charge transfer at the interface from the iridate to the manganite side discussed in part 3 of the paper is the main reason for ferromagnetic interaction in the iridate/manganite heterostructure. The electrons transferred to the manganite add ferromagnetic ordering through the double exchange interaction, while the iridate part becomes ferromagnetic due to the doping of the half-filled Mott-Hubbard insulator [15, 17, 31]. The occurrence of magnetism at the interface caused by hybridization of Mn and Ir orbitals is the reason for the axis of easy magnetization rotation between the crystallographic directions: (110) $La_{0.7}Sr_{0.3}MnO_3$ and (001) $La_{0.7}Sr_{0.3}MnO_3$ in manganite/iridate superlatixe [15, 31].

Thus in PNR experiment we observed turn of magnetic vector at remanence on 26 degree (Fig. 6b). Similar turn of easy axis direction was observed recently in LaNiO$_3$/DyScO$_3$ superlattice [51] and was explained by appearance of magnetic moment at Dy with strong anisotropy non-collinear to the easy axis of nickelate. Strong exchange interaction of Ni and Dy atoms at the interface leads to the turn of magnetic moment in nickelate towards easy axis of DyScO$_3$ layer.

5. Magnetic anisotropy in heterostructure

To record the magnetic resonance spectra, we used the Bruker ER 200 spectrometer operating in the X-band (f = 9.6 GHz) with the Oxford cryogenic ESR 900 insert. The external magnetic field laid in the sample plane for all FMR experiments and the magnetic component of the microwave field was directed along the normal to the substrate [52]. All spectra were taken on samples with dimensions 2.5x2.5 mm$^2$ with thicknesses 10 nm and 4nm for SIO and LSMO correspondingly. Note the thickness of the both LSMO and SIO were 4-5 times less then for heterosructure in neutron experiment and the LSMO film is a bottom layer.

Figure 7a shows the ferromagnetic resonance (FMR) spectra dP/dH (here P is an absorption value) of an autonomous LSMO film T=300K and for heterostructures: SIO/LSMO at two temperatures T =90 and 40 K. Note that at room temperature only the FMR line from the LSMO layer is observed, since the sensitivity of the spectrometer does not allow recording the absorption spectrum from the paramagnetic SIO layer. It can be seen from the Fig.7a that deposition of an SIO layer broadens the FMR line. The observed broadening can be caused by additional channels of magnetization relaxation due to the leakage of magnetization across the SIO/LSMO interface due to the spin current.

Figure 7b shows the resonance field and liwidth of FMR spectrum for the heterostructure structures, obtained under the condition that the external magnetic field was directed along the difficult axis of the uniaxial magnetic anisotropy. This direction of the external magnetic field was chosen from the condition of the minimum contribution of magnetic anisotropy to the resonance relation for FMR [52]. Thus, we can assume that the obtained temperature dependence characterizes the change in the magnetization of the heterostructure. In this approximation, we can say that a decrease in the resonance field is due to an increase in the magnetization of the sample. It follows from Fig.7b that the Curie temperature for the heterostructure is higher than 340 K, which is typical for an autonomous LSMO film on an NGO substrate. At the same time, a sharp decrease in the H$_0$ value in both structures with a decrease in temperature below 50 K. It is impossible to explain such a drop by a sharp increase in the magnetization of the LSMO layer, since this contradicts the measurements of autonomous LSMO films[52]. Typically, with decreasing temperature, the dependence $H_0$ (T) saturates at temperatures below 100 K.

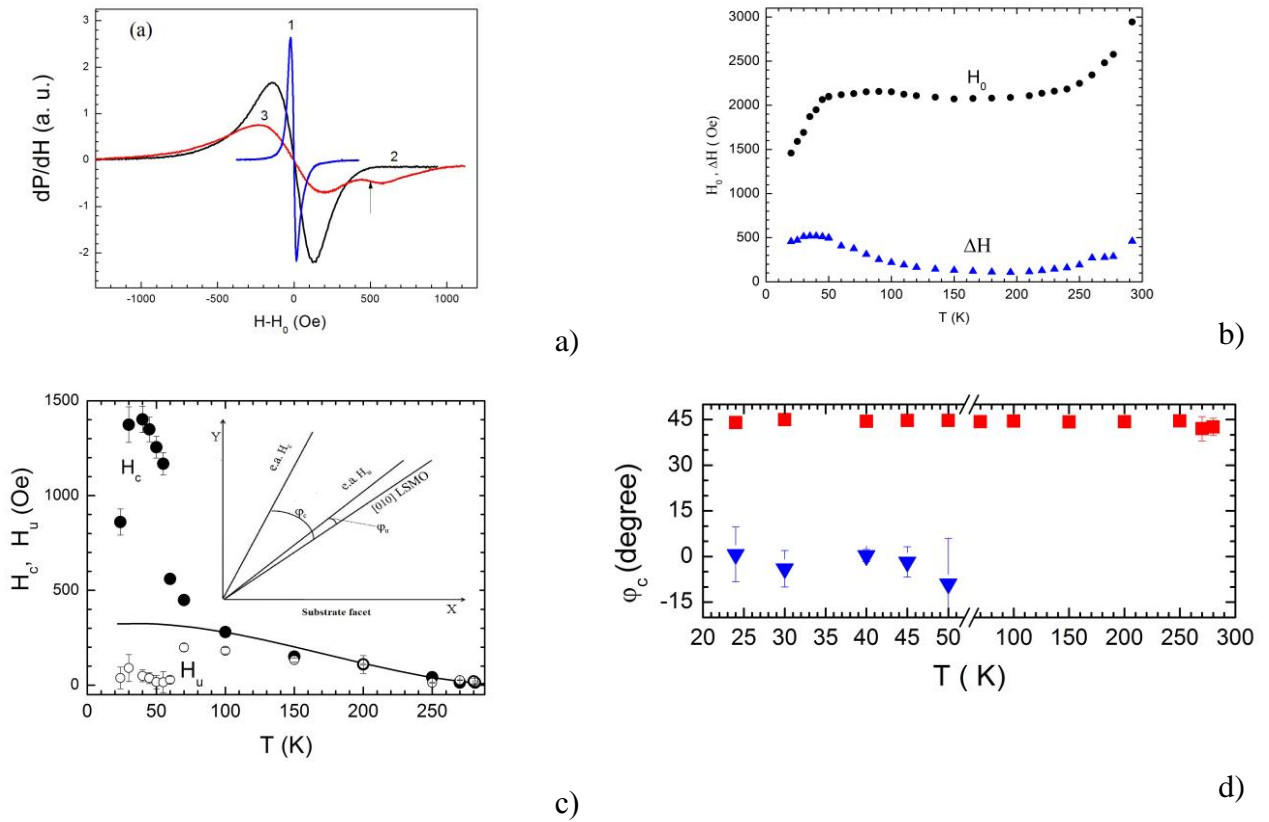

Fig.7 a) Ferromagnetic resonance spectra dP/dH(H) for LSMO film and SIO/LSMO heterostructures: (1)-15 nm thick stand alone LSMO film at T=300K and SIO/LSMO heterostructures with 12nm thicknesses of LSMO and 10 nm SIO for (2)-T = 90K and (3)-T=40K. Additional line on FMR spectrum is indicated by arrow. b) The temperature dependences of resonance field $H_0$ and linewidth $\Delta H$ for heterostructure SIO/LSMO. c) Temperature dependence of biaxial magnetic anisotropy $H_c$ (filled circles) and uniaxial magnetic anisotropy $H_u$ (open circles) for the heterostructure, solid line is the temperature dependence of $H_c$ value, averaged over various film structures[55], which included the LSMO layer (see text). Inset show the typical position of easy axis $H_c$ and $H_u$ at room temperature. d) angle of biaxial magnetic anisotropy $\varphi_c$ for the main line FMR (squares) and for an additional line of FMR (inverted triangles).

LSMO films grown on NGO substrates possess an induced planar uniaxial anisotropy reaching hundreds of Oe at room temperature in addition to the biaxial one inherent in the cubic structure of LSMO film [52-55]. To determine the magnetic anisotropy of the heterostructure, the angular dependences of the FMR spectrum were recorded at fixed temperatures (see Appendix). The value of the resonant field is a function of the magnitude of the equilibrium magnetization $M_0$ and anisotropy fields $H_u = 2K_u/M_0$, $H_c = 2K_c/M_0$, where $K_u$ and $K_c$ are the uniaxial and the biaxial cubic anisotropy constants correspondingly. The temperature dependences of the uniaxial $H_u$ and biaxial magnetic anisotropy $H_c$ for SIO/LSMO heterostrucuture are shown in Fig. 7c. At high (room) temperature, $H_u$ is higher than $H_c$. With a decrease in temperature, $H_c$ dominates in the heterostructure, despite the small increase in $H_u(T)$ was observed (Fig.7c). Earlier it was shown that the temperature dependences of the biaxial cubic magnetic anisotropy of LSMO films are practically independent on the layers in the heterostructure with LSMO and the substrates on which they were grown [54-56]. We observed the same behavior of $H_c(T)$ at high temperatures T>100K (see Fig. 7c). In SIO/LSMO heterostructures, there is an additional increase in the growth of $H_c$ (T) at T <50 K, which is absent upon the contact of 3d oxides [55]. A similar increase in the magnetization of the LSMO layer at temperatures below 150 K (Curie temperature for SRO) was observed in the LSMO/SRO heterostructure (here SRO is $SrRuO_3$) and interpreted as the appearance of interlayer exchange interaction after the SRO layer passed into ferromagnetic state [55]. Note, that the turning of the heterostructure SIO/LSMO magnetization vector of closer to the external field was observed in neutron experiment (see part 4).

The relationship between the magnitude of the magnetic anisotropy and the deformation of the crystal lattice is a well-known fact, both for single-layer films and for multilayer film structures [55, 56]. In turn, the magnetocrystalline anisotropy is determined by the spin-orbit coupling (SOI) and the anisotropy of the structure, which includes the distances and angles of the magnetic atom with the oxygen atom. In [15], superlattices consisting of repeatedly repeated LSMO / SIO pairs were investigated in detail. It was shown that the interaction of the LSMO and SIO layers through the Ir – O – Mn bond in the direction perpendicular to the plane of the boundary plays an important role in increasing the magnetic anisotropy. The slightest deformations and rotations of this bond effectively influence the exchange interactions across the interface and, under certain combinations, can create conditions for sufficiently strong exchange interactions in the entire iridate layer.

In the temperature range, where a sharp increase in $H_c$ (T) is observed at T <70 K (see Fig.8b), an additional FMR line appeared, which indicates the appearance of an additional ordered ferromagnetic state in the heterostructure. Signs of ferromagnetism at the SIO/LSMO interface were recorded in [16, 32, 54, 56]. The emergence of a new FMR line can be caused either by the appearance of a ferromagnetic order in the SIO film, as observed in the SIO/LSMO [15] superlattices and predicted theoretically in [40], or from the spins of the LSMO layer, which become ferromagnetic at a lower temperatures. The transferred charge (see part 2) plays an important role in altering the magnetic interactions near the interface. The density-functional results [42] show that the interfacial magnetism is controlled by a net charge transfer at the interface from the SIO to the manganite. The doped electrons turn the manganite part metallic and change ferromagnetic states via the double exchange mechanism. The hole doped iridate part, on the other hand, behaves like a half-filled Mott-Hubbard insulator and becomes ferromagnetic [42]. The emergence of ferromagnetism at the interface of the 3d manganite and 5d iridate interface is in agreement with the experimental observation [16] and unravel its mechanism.

Figure 7d shows the temperature dependences of the directions of the axes of easy magnetization of biaxial magnetic anisotropy $\varphi_c$ for the main and for the additional FMR lines. Here the angles $\varphi_c$ are taken relative to the direction [010] LSMO. It can be seen that in the region of existence of an additional ferromagnetic spin system, the easy axes of the uniaxial magnetic anisotropy of the two spin systems are rotated by 90º, and the axes of the biaxial magnetic anisotropy of these systems are rotated by 45º. Thus, it can be argued that the additional FMR line cannot belong to the spins of the epitaxially grown LSMO layer. The emergence of ferromagnetism in a thin SIO layer was observed in various superlattices consisting of SIO layers: (SrMnO3/SIO) n [16, 17], (SrTiO3/SIO) n [14], (LSMO/SIO) n [15, 18]. It can be assumed that the SIO film enters a ferromagnetic state near the temperature of 50 K [34]. The interlayer exchange interaction of two ferromagnets causes a sharp decrease in the resonant field. The exact value of the magnetization of the LSMO film, taking into account the effect of magnetic anisotropy, can be determined from measurements of the angular dependences of the FMR spectra at different temperatures [56].

6. Spin current

The width of FMR for SIO/LSMO heterostructure exceeds the width one for the single LSMO film (see Fig. 7a). A possible reason for the broadening of the FMR line of the heterostructure is additional spin relaxation due to spin current in the structure ferromagnetic/normal metal in FMR. The spin pumping mechanism creates a spin current from a ferromagnetic to a normal metal. As a result, an additional channel of torque outflow from the ferromagnetic is formed, which leads to an increase of the damping parameter [57-60]. Using the following expressions, we calculated the spin mixing conductance in heterostructure SIO/LSMO

$$g_{eff}^{\uparrow\downarrow} = \frac{4\pi\gamma_g M_s t_{LSMO}}{g\mu_\beta \omega_f}\left(\Delta H_{SIO/LSMO} - \Delta H_{SIO}\right) \tag{1}$$

For the following parameters: $M_s$= 300 Oe is the magnetization LSMO film, $t_{LSMO}$= 12 nm is the thickness for LSMO film, $\mu_B$=9.274•$10^{-21}$ erg/G is the Bohr magneton, g=2 is Lande factor, $\gamma_g$=17,605•$10^6$ $s^{-1}G^{-1}$ is gyromagnetic ratio for free electron $\omega_f$=2π•9.51•$10^9$ $s^{-1}$ is the microwave angular frequency in our case. At the room temperature we got $\Delta H_{SIO/LSMO}$ -$\Delta H_{LSMO}$ =20 Oe, $g_{eff}^{\uparrow\downarrow}$= 0.95·$10^{18}$ $m^{-2}$ for SIO/LSMO heterostructures. For comparison with interfaces $SrRuO_3/La_{2/3}Sr_{1/3}MnO_3$, $Pt/Ni_{80}Fe_{20}$ and Pt/YIG (YIG is yttrium iron garnet) were obtained for $g_{eff}^{\uparrow\downarrow}$ values 5•$10^{18}$ $m^{-2}$ [61], 2.1·$10^{19}$ $m^{-2}$ [57] and 4.8·$10^{20}$ $m^{-2}$ [58] correspondingly.

For the detection of spin current we used the method based on inverse spin Hall effect (ISHE) [62]. In the material with strong spin-orbit interaction a spin current gives rise to an electric current. The relation between spin and electric currents is determined by the dimensionless spin Hall angle $\theta_{SH}$:

$$\vec{j}_{ISHE} = \theta_{SH} \frac{e}{\hbar}\left[\vec{n} \times \vec{j}_s^0\right] \quad (2)$$

where $\vec{n}$ is a unit vector in the direction of the spin momentum flow.

The sample is a strip of heterostructure SIO/LSMO on NGO substrate with electric silver contacts at the edges for the voltage measurements (see Fig. 8b). It was placed in the central plane of rectangular $TE_{01}$ microwave cavity. The microwave pumping was produced by Gunn diode with output up to 130 mW and at the frequency $\omega_f/2\pi = 9.0$ GHz. The signal V(H) was accumulated by sweeping the external field H across the resonance value $H_0$. By rotating the external field H in the film plane we measured the voltage for angles in the range 180 degrees with 10 degrees step.

The typical signal V(H) detected at the SIO/LSMO heterostructure is shown at Fig. 8a. Spin current generated under ferromagnetic resonance condition is converted into electric current due to ISHE. In turn, this current creates the dc voltage. In addition, one should take into account dc voltage arising due to the presence of anisotropic magnetoresistance (AMR) in LSMO layer. As a result, the full dc voltage and the angle dependence should be written as the following [63]:

a)

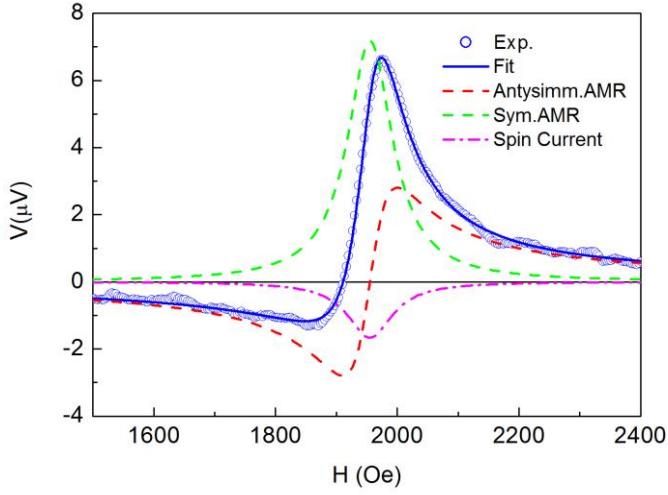
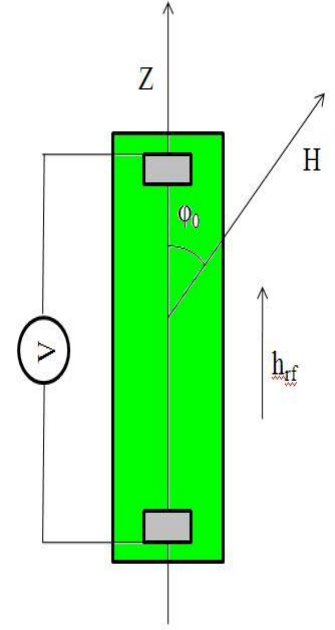

b)

Fig.8. a) The voltage arising in the SIO/LSMO heterostructure during the magnetic field sweeping at room temperature (blue points). The dashed green and red lines represent symmetric and antisymmetric part of the AMR signal correspondingly and magenta line represents signal from the spin pumping. Solid blue line is the sum of three contributions (see Eq. 3). b) topology for voltage measuring.

$$V(H,\varphi_0) = V_{SP} \cdot L(H) \cdot \sin^3 \varphi_0 + [V_{AMR}^{Sym} \cdot L(H) + V_{AMR}^{Antisym} \cdot L'(H)] \cdot \sin 2\varphi_0 \cdot \sin \varphi_0 \qquad (3)$$

Here the $L(H) = (\Delta H)^2/[(H - H_0)^2 + (\Delta H)^2]$ is the symmetric Lorentzian function with the resonance field $H_0$ and half-width $\Delta H$ of FMR spectrum. $L'(H) = \Delta H(H - H_0)/[(H - H_0)^2 + (\Delta H)^2]$ is an antisymmetric function and $\varphi_0$ is the angle between the directions of the external magnetic field $H$ (the magnetization is parallel of $H$ in our case) and charge current [60, 64]. The voltage from the spin pumping and AMR changes sign upon inversion of the $H$ direction. Since parasitic contribution is constant for the opposite orientations of the magnetic field we use the difference for the signals with opposite orientations of the magnetic field in order to exclude this parasitic contribution.

In order to divide the symmetric signal into the effects of spin pumping and anisotropic magnetoresistance we measured the voltage dependence upon the angle $\varphi_0$. By fitting the angular dependence with formula (3) we obtained the following ratio between amplitude of the symmetric signal of anisotropic magnetoresistance and signal of spin pumping: $V_{SP}/V_{AMR}^{Sym} = 0.23 \pm 0.03$.

7. Conclusion

The measurements of the dc transport and magnetic properties of $SrIrO_3/La_{0.7}Sr_{0.3}MnO_3$ epitaxial heterostructures showed the presence of unusual properties of the interface between the materials. The dc measurement indicates the presence of a conduction channel at the iridate/manganite interface. The magnetization and magnetic profile of SIO/LSMO heterostructure was investigated by SQIUD and neutron scattering. Comparison of FMR linewidth obtained for the LSMO film and iridate heterostructure with SIO on top LSMO film the spin mixing conductance was determined. At the temperatures below 60 K a sharp increase of biaxial anisotropy ($H_c$) was observed and an additional FMR line appeared, which indicates on the appearance of an additional ordered ferromagnetic state in the heterostructure. One of the possible causes may be the appearance of ferromagnetism in the paramagnetic SIO film near interface SIO/LSMO. The neutron scattering data can be explained by the turn of magnetization vector of the heterostructure (mainly LSMO) on 26 degree closer to the external field. We have measured the dc voltage on the SIO film caused by spin pumping and by the anisotropic magnetoresistance in the heterostructure in presence of FMR in the heterostrcuture.

Acknowledgments

V.A. Atsarkin, K.I. Constantinian, B. Kaimer, T. Keller, A.A. Klimov, A.M. Petrzhik and A.V. Shadrin are acknowledged for experimental help and fruitful discussion. This work was partially supported by Russian Foundation for Basic Research, projects No. 18-37-00170 and No. 19-07-00143

References

1.	Y.Tokura, Phys. Today **56**, 50 (2003).

2.	E.Dagotto, Science **309**, 257 ( 2005).

3.	S.J. Moon, H. Jin, K.W. Kim, W.S. Choi, Y.S. Lee, J. Yu, G. Cao, A. Sumi, H. Funakubo, C. Bernhard, T.W. Noh, Phys. Rev. Lett. **101**, 226402 (2008).

4.	D. Pesin, L. Balents, NaturePhysics **6**, 376 (2010).

5.	W. Witczak-Krempa, G. Chen, Y. B. Kim, and L. Balents, Annu. Rev. Condens. Matter. Phys. **5**, 57 (2014).

6.	R. Schaffer, E. Lee, B. Yang and Y. Kim, Rep. Prog. Phys. **79**, 094504 (2016).

7.	A. Shitade, H. Katsura, J. Kuneš, X.-L. Qi, S.-C. Zhang, N. Nagaosa, Phys. Rev. Lett. **102**, 256403 (2009).

8.	D. Xiao, W. Zhu, Y. Ran, N. Nagaosa, S. Okamoto, Nature Commun. **2**, 596 (2011).

9. F. Wang, Y. Ran, Phys. Rev. B **84**, 241103 (2011).

10. X. Wan, A.M. Turner, A. Vishwanath, S.Y. Savrasov, Phys. Rev. B, **83**, 205101 (2011).

11. J. G. Kim, D. Casa, M.H. Upton, T. Gog, Y.-J. Kim, J.F. Mitchell, M. van Veenendaal, M. Daghofer, J. van den Brink, G. Khaliullin, B.J. Kim, Phys. Rev. Lett. **108**, 177003 (2012).

12. Y. K. Kim, N.H. Sung, J.D. Denlinger and B.J. Kim, Nature Physics. **12,** 37 (2016).

13. I. Fina, X. Marti, D. Yi, J. Liu, J.H. Chu, C. Rayan-Serrao, S. Suresha, A.B. Shick, J. Zˇelezny, T. Jungwirth, J. Fontcuberta & R. Ramesh, Nature Communication **5,**4671 (2014).

14. J. Matsuno, K. Ihara, S. Yamamura, H. Wadati, K. Ishii, V. V. Shankar, H.-Y. Kee and H. Takagi, *Phys. Rev. Lett.***114** 247209 (2015)  J. Matsuno, N. Ogawa, K. Yasuda, F.Kagawa, W. Koshibae, N. Nagaosa,Y. Tokura and M.Kawasaki,  Sci.Adv. **2** 1600304 (2016).

15. Di Yi, Jian Liu, Shang-Lin Hsu, Lipeng Zhang, Yongseong Choi, Jong-Woo Kim, Zuhuang Chen, James D. Clarkson, Claudy R. Serrao, Elke Arenholz, Philip J. Ryan, Haixuan Xu, Robert J. Birgeneau, and Ramamoorthy Ramesh,  Proc. Nat Acad. Sci. **113** 6397 (2016).

16. John Nichols, Xiang Gao, Shinbuhm Lee, Tricia L. Meyer, John W. Freeland, Valeria Lauter, Di Yi, Jian Liu, Daniel Haskel, Jonathan R. Petrie, Er-Jia Guo, Andreas Herklotz, Dongkyu Lee, Thomas Z. Ward, Gyula Eres, Michael R. Fitzsimmons,  and  Ho Nyung Lee, Nature Communications **7**, 12721 (2016).

17. S. Okamoto, J. Nichols,C. Sohn,S.Y. Kim, T.W. Noh and H.N. Lee, Nano Lett. **17** 2126, (2017).

18.  Rachna Chaurasia, K.C. Kharkwal, A.K. Pramanik,  Physics Letters A **383** 1642 (2019).

19. J.M. Longo, J.A. Kafalas, R.J. Arnott, J. Solid State Chem. **3**, 174(1971).

20. Fei-Xiang Wu, Z. Jian, L.Y. Zhang, Y.B. Chen, Z. Shan-Tao, G. Zheng-Bin, Y. Shu-HuaY and C. Fan-Feng, J. Phys.Condens. Matter **25,** 125604 (2013).

21. A. Biswas, Y.H. Jeong, Current Applied Physics **17**, 605 (2017).

22. Lunyong Zhang, Qifeng Liang, Ye Xiong, Binbin Zhang, Lei Gao, Handong Li, Y. B. Chen, Jian Zhou, Shan-Tao Zhang, Zheng-Bin Gu, Shu-hua Yao, Zhiming Wang, Yuan Lin, and Yan-Feng Chen*,* Phys. Rev. B**91,** 035110 (2015).

23. Y.F. Nie, P.D.C. King, C.H. Kim, M. Uchida, H.I. Wei,1 B.D. Faeth,1 J.P. Ruf, J.P.C. Ru_, L. Xie, X. Pan, C.J. Fennie, D.G. Schlom, and K.M. Shen, Phys. Rev. Lett. **114,** 016401 (2015).

24. Lunyong Zhang, Bin Pang,Y. B Chen &Yanfeng Chen, Critical Reviews in Solid State and Materials Sciences  **43**,  367 (2018).

25. Yuxue Liu, Hiroshi Masumoto and Takashi Goto, Materials Transactions,  **46**, 100 (2005).

26. L. Fruchter, O. Schneegans, and Z. Z. Li, J. Appl. Phys. **120**, 075307 (2016).

27. J.H. Gruenewald, J. Nichols, J. Terzic, G. Cao, J.W. Brill, S.S.A. Seo, J. Mater. Res. **29,** 2491 (2014)

28. Yu.V. Kislinskii, G.A. Ovsyannikov, A.M. Petrzhik, K.Y. Constatinian, N.V. Andreev, T.A. Sviridova, Physics of the Solid States, **57**, 2519 (2015).

29. B.J. Kim , Hosub Jin, S. J. Moon, J.-Y. Kim, B.-G. Park, C. S. Leem, Jaejun Yu, T. W. Noh, C. Kim, S.-J. Oh, J.-H. Park, V. Durairaj, G. Cao, and E. Rotenberg, Phys.Rev. Lett. **101,** 076402 (2008).

30. J. G. Zhao, L. X. Yang, Y. Yu, F. Y. Li, R. C. Yu, Z. Fang, L. C. Chen, and C. Q. Jin, J. Appl. Phys. **103**, 103706 (2008).

31. D. Yi, Charles L. Flint, Purnima P. Balakrishnan, Krishnamurthy Mahalingam, Brittany Urwin, Arturas Vailionis, Alpha T. N'Diaye, Padraic Shafer, Elke Arenholz, Yongseong Choi, Kevin H. Stone, Jiun-Haw Chu, Brandon M. Howe, Jian Liu, Ian R. Fisher, and Yuri Suzuki, Phys. Rev. Lett. **119,** 077201 (2017).

32. T. A. Shaikhulov, G.A. Ovsyannikov, V.V. Demidov and N. V. Andreev, Journal of Experimental and Theoretical Physics **129**, 112 (2019).

33. S. Crossley , A. G. Swartz, K. Nishio, Y. Hikita, and H. Y. Hwang, Physical Review **100**, 115163 (2019).

34. V.V. Demidov, N.V. Andreev, T.A. Shaikhulov, G.A. Ovsyannikov, Journal of Magnetism and Magnetic Materials **497** 165979 (2020).

35. G. A. Ovsyannikov, A. M. Petrzhik, , I. . Borisenko, A. A. Klimov, Yu. A. Ignatov, V. V. Demidov and S. A. Nikitov, Journal of Experimental and Theoretical Physics **108**, 48 (2009).

36. H.C. Montgomery, Journal of Applied Physics **42**, 2971 (1971).

37. A. Urushibara, Y. Moritomo, T. Arima, A. Asamitsu, G. Kido, and Y. Tokura, Phys. Rev. B **51**, 14103 (1995).

38. E. J. Moon, A. F. May, P. Shafer, E. Arenholz, and S. J. May, Phys. Rev.B **95**, 155135 (2017).

39. C.T. Boone, J.M. Shaw, H.T.Nembach and T.J. Silva, Journal of Applied Physics **117**, 223910 (2015).

40. S. Thiel, G. Hammer, A. Schmehl, C. W. Schneider, J. Mannhart, Science, **313**, 1942 (2006).

41. M.Ye. Zhuravlev, A.V. Vedyayev, M.S. Titova , N.V. Ryzhanova , D. Gusakova, Journal of Magnetism and Magnetic Materials **441,** 572 (2017).

42. Sayantika Bhowal, and Sashi Satpathy, AIP Conference Proceedings **2005**, 020007 (2018); Phys. Rev. B**99**, 245145 (2019).

43.	A. Ohtomo and H. Hwang, Nature (London) **427**, 423 (2004).

44.	C. Kittel, Introduction to Solid State Physics (Wiley,New York, 1996), 7th ed..

45.	Yu. N. Khaydukov, B. Nagy, J.-H. Kim, T. Keller, A.Ruhm, Yu.V. Nikitenko, K.N. Zherenkov, J. Stahn, L.F. Kiss, A. Csik, L. Bottyan, A.L. Aksenov, Journal of Experimental and Theoretical Physics Lett, **98**, 107, (2013).

46.	H. Zabel, R.K. Theis Bohl, Toperver Handbook of Magnetism and Advanced Magnetic Material (Wiley, 2007).

47.	Yu. N. Khaydukov, G. A. Ovsyannikov, A. E. Sheyerman, K. Y. Constantinian, L. Mustafa, T. Keller, M. A. Uribe-Laverde, Yu. V. Kislinskii, A. V. Shadrin, A. Kalaboukhov, B. Keimer, and D. Winkler , Phys. Rev. B **90**, 035130 (2014).

48.	Yu. Khaydukov, A. M. Petrzhik, I. V. Borisenko, A. Kalabukhov, D. Winkler, T. Keller, G. A. Ovsyannikov, and B. Keimer, Phys. Rev. B **96,** 165414 (2017).

49.	G. A. Ovsyannikov, A. E. Sheyerman, A. V. Shadrin,Yu. V. Kislinskii, K. Y. Constantinian, and A. Kalabukhov, Journal of Experimental and Theoretical Physics Lett. **97**, 145 (2013).

50.	Yu. Khaydukov, Yu.N. Nikitenko, Nucl. Instrum. Methods Res.A **629,** 245 (2011).

51.	M. Bluschke, A. Frano, E. Schierle, M. Minola, M. Hepting, G. Christiani,G. Logvenov, E. Weschke, E. Benckiser, and B. Keimer, Phys Rev Lett **118**, 207203 (2017).

52.	V. V. Demidov, I. V. Borisenko, A. A. Klimov, G. A. Ovsyannikov, A. M. Petrzhik, S. A. Nikitov, Journal of Experimental and Theoretical Physics **112**, 825 (2011).

53.	Y. Moritomo, A. Asamitsu, and Y. Tokura, Phys. Rev. B **51**, 16491 (1995).

54.	V.V. Demidov, G.A. Ovsyannikov, A.M. Petrzhik, I.V. Borisenko, A.V. Shadrin, and R. Gunnarsson, J. Appl. Phys. **113,** 163909 (2013).

55.	V. V. Demidov and G. A. Ovsyannikov, J. Appl Phys. 122, 013902 (2017).

56.	V.V. Demidov, , N.V. Andreev, T.A. Shaikhulov, G.A. Ovsyannikov, Journal of Magnetism and Magnetic Materials, 497, 165979 (2020) .

57.	O. Mosendz, V. Vlaminck, J.E. Pearson, F.Y. Fradin, G. E.W. Bauer, S.D. Bader, and A. Hoffmann, Phys. Rev. **82**, 214403 (2010).

58.	M. Rezende, R.L. Rodrıguez-Suarez, M.M. Soares,L. H. Vilela-Le, D. Ley Dom_ınguez, and A. Azeved, Appl. Phys. Lett. **102**, 012402 (2013).

59.	T.G.A. Verhagen, H.N. Tinkey, H.C. Overweg, M. van Son, M. Huber, J.M. van Ruitenbeek and J. Aarts, J. Phys. Condens. Matter **28,** 056004 (2016) (10pp).

60.	A. Azevedo, L.H. Vilela-Leao,R. L. Rodrıguez-Suarez, A.F. Lacerda Santos, and S. M. Rezende, Phys.Rev. B**83**, 144402 (2011).

61. S. Emori, U.S. Alaan, M.T. Gray, V.Sluka, Y. Chen, A.D.Kent, Y.Suzuki, Phys. Rev **B94**, 224423 (2016).

62. V.A. Atsarkin, I.V. Borisenko, V.V. Demidov and T.A. Shaikhulov, J. Phys. D: Appl. Phys. **51** 245002 (2018)

63. V. A. Atsarkin and B. V. Sorokin, Journal of Experimental and Theoretical Physics **119** 567 (2014).

64. S. Emori, U.S. Alaan, M.T. Gray, V.Sluka, Y. Chen, A.D.Kent, Y.Suzuki, Phys. Rev **B94**, 224423 (2016).

65. T. M. Vasilevskaya and D. I. Sementsov, Phys. Met. Metallography **108**, 321 (2009).

Appendix

The method for determining the parameters of the magnetic anisotropy consisted in processing the angular dependences of the resonant fields of the FMR spectra. The solution of the linearized Landau-Lifshitz-Gilbert equation is used for the evolution of the magnetization $M$ in an external constant magnetic field $H$ under the action of the magnetic component of the radio-frequency field. This solution gives an analytic connection between the external resonance field $H_0$ and the frequency $\omega$ under FMR conditions [52, 63.].

$$\left(\frac{\omega}{\gamma}\right)^2 = \left(4\pi M_0 + H_0(\varphi) + H_u \cos^2 \varphi_u + H_c \frac{1+\cos^2 2\varphi_c}{2}\right)\left(H_0 + H_u \cos 2\varphi_u + H_c \cos 4\varphi_c\right) \quad (1a)$$

Here $\gamma$ is the gyromagnetic ratio, $M_0$ is the equilibrium magnetization, $H_u = 2K_u/M_0$, $H_c = 2K_c/M_0$, $K_u$ is the uniaxial anisotropy constant, and $K_c$ is the biaxial cubic anisotropy constant. As a result, the values of $K_u$, $K_c$, $M_0$, as well as the angles between the easy axis of the uniaxial anisotropy and the external magnetic field $\varphi_u$ and between the easy axis of the biaxial cubic anisotropy and the external magnetic field $\varphi_c$ are determined from the angular dependence of the magnitude of the resonant magnetic field $H_0$. Both easy axes lie in the plane of the substrate.

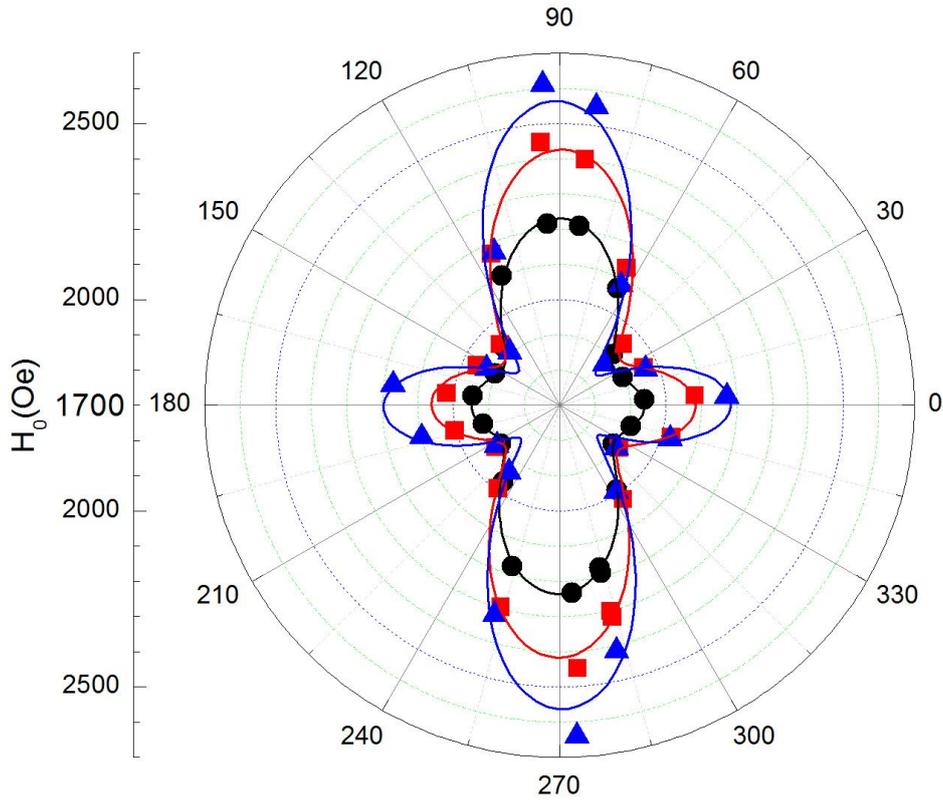

Fig.1ap. Angular dependences of the resonance field for the SIO/LSMO heterostructure obtained at temperatures: T=150K (filled circles), T=90K (red squares), T=40 (blue triangular). Color solid lines are the fitting according to formula (1) with the parameters indicated in table 1a

Fig. 1ap shows the $H_0(\varphi)$ for the SIO/LSMO heterostructure at room temperature. The external magnetic field is rotated around the normal to the film plane by an angle $\varphi$. The angle was measured from the direction of [010]LSMO. The external magnetic field and the magnetic component of the microwave field were in the plane of the film. The change of resonant field when the angle changes is due to the planar magnetic anisotropy of the SIO/LSMO heterostructure. The angular dependence was described by a resonance relation (2) taking into account magnetic uniaxial and cubic plane anisotropies [52].

Table 1ap.

Magnetic parameters of the heterostructure SIO/LSMO, obtained from angle dependence of resonance field.

| T, K | $M_0$, Oe | $H_u$, Oe | $H_c$, Oe | $\varphi_u$, grad | $\varphi_c$, grad |
|---|---|---|---|---|---|
| 150 | 309.5±0.6 | 165±3 | 105±3 | 1.2±0.6 | 44.9±0.5 |
| 90 | 277±2 | 200±8 | 199±9 | 1±1.3 | 45.1±0.7 |
| 40 | 260±3 | 215±17 | 325±19 | 2±2.5 | 44.0±0.9 |